**Assessing The Spatially Heterogeneous Transportation Impacts Of Recurrent Flooding In The Hampton Roads Region – Part 1: Auto Accessibility**

**Luwei Zeng**
Graduate Research Assistant
Department of Engineering System and Environment
University of Virginia, Charlottesville, VA, 22903
lz6ct@virginia.edu

**T. Donna Chen, PE, Ph.D.**
Corresponding Author
Assistant Professor
Department of Engineering Systems and Environment
University of Virginia, Charlottesville, VA 22903
tdchen@virginia.edu

**John S. Miller, PE, Ph.D.**
Research Scientist
Virginia Transportation Research Council
Charlottesville, VA, 22903
John.Miller@VDOT.Virginia.gov

**Jonathan L. Goodall**
Professor
Department of Engineering, Systems and Environment
University of Virginia, Charlottesville, VA 22903
goodall@virginia.edu

**Faria Tuz Zahura**
Graduate Research Assistant
Department of Engineering Systems and Environment
University of Virginia, Charlottesville, VA 22903
Fz7xb@vriginia.edu

**ABSTRACT**

Recurrent flooding has increased rapidly in coastal regions due to sea level rise and climate change. A key metric for evaluating transportation system degradation is accessibility, yet the lack of temporally and spatially disaggregate data means that the impact of recurrent flooding on accessibility—and hence transportation system performance—is not well understood. Using crowdsourced WAZE flood incident data from the Hampton Roads region in Virginia, this study (Part 1) examines changes in the roadway network accessibility for travelers residing in 1,113 traffic analysis zones (TAZs) across five time-of-day periods. Additionally, a social vulnerability index framework is developed to understand the socioeconomic characteristics of TAZs that experience high accessibility reduction under recurrent flooding.

Results show that TAZs experience the most accessibility reduction under recurrent flooding during the morning peak period (6 to 9am) with large differences across different zones, ranging from 0% to 49.6% for work trips (with population-weighted mean reduction of 1.71%) and 0% to 87.9% for non-work trips (with population-weighted mean reduction of 0.81%). Furthermore, the social vulnerability analysis showed that zones with higher percentages of lower socio-economic status, unemployed, less educated, and limited English proficiency residents experience greater accessibility reduction for work trips. In contrast to previous studies that aggregate the effects of recurrent flooding across a city, these results demonstrate that there exists large spatial and temporal variation in recurrent flooding's impacts on accessibility. This study also highlights the need to include social vulnerability analysis in assessing impacts of climate events, to ensure equitable outcomes as investments are made to create resilient transportation infrastructure.

## 1. INTRODUCTION

Accessibility—the ease with which one can attain a social or economic goal through the transportation and land development system—is an important measure of effectiveness for the transportation system. Wachs (*1*) defined the relationship between the spatial distribution and intensity of development, and the quantity and quality of travel within a region, as the region's accessibility. This ease of reaching opportunities through the transportation network (*2*) is thus the combination of mobility offered by the network to potential destinations and the value of destinations (*3*). For a given location, disruptions to either the transportation network or to desired destinations will be reflected in that location's aggregate accessibility.

One increasingly common disruption to the transportation network in coastal regions is recurrent flooding, which is exacerbated by accelerated sea level rise (*4*). Recurrent flooding can be caused by storm surges or tide events such that it occurs repeatedly at the same locations (*5*). Recurrent flooding has not received as much attention in research as large-scale flooding under extreme events (*6*), since recurrent flooding typically increases travel impedance without eliminating access. However, such recurrent flooding materially affects transportation system performance (*7–9*), such that the cumulative impact is non-negligible and has been increasing: an almost five-fold increase in Washington, D.C. was reported when comparing average annual recurrent flooding from 1930-1970 to the "last two decades" (*6*). This increase and its deleterious impacts are expected to continue: Jacobs et. al. in 2018 estimated the total traffic delay along the US eastern coast caused by recurrent flooding to be over 1.2 billion vehicle-hours by 2060 (*10*). Because elements of the transportation system and the built environment may have varying degrees of resistance to recurrent flooding, it is possible that such flooding will have disparate impacts on populations with different socioeconomic characteristics (*11*, *12*). Moreover, the cumulative impact of these diffused, low-cost flooding events may be comparable to property value losses associated with less frequent but larger scale weather events (*6*). Because equity impacts are fundamental to transportation investment decisions, there is a need to determine the extent to which recurrent flooding impacts vary within a region.

This paper (Part 1) examines the spatially and temporally heterogeneous impacts of recurrent flooding on auto accessibility throughout the coastal Hampton Roads region in Virginia, USA. The research objectives are to (1) develop a methodology for quantifying accessibility impacts of recurrent flooding with easily accessible crowdsourced data, (2) demonstrate this methodology on a coastal case study region, and (3) identify locations and populations whose accessibility is most impacted by recurrent flooding. This case study focuses on the Hampton Roads region in the mid-Atlantic portion of the U.S.—a region of roughly 3,700 mi$^2$ with 1.7 million people and 1 million jobs (Figure 1). A second and parallel paper (Part 2) investigates the accessibility of the transit network under recurrent flooding in the same study area.

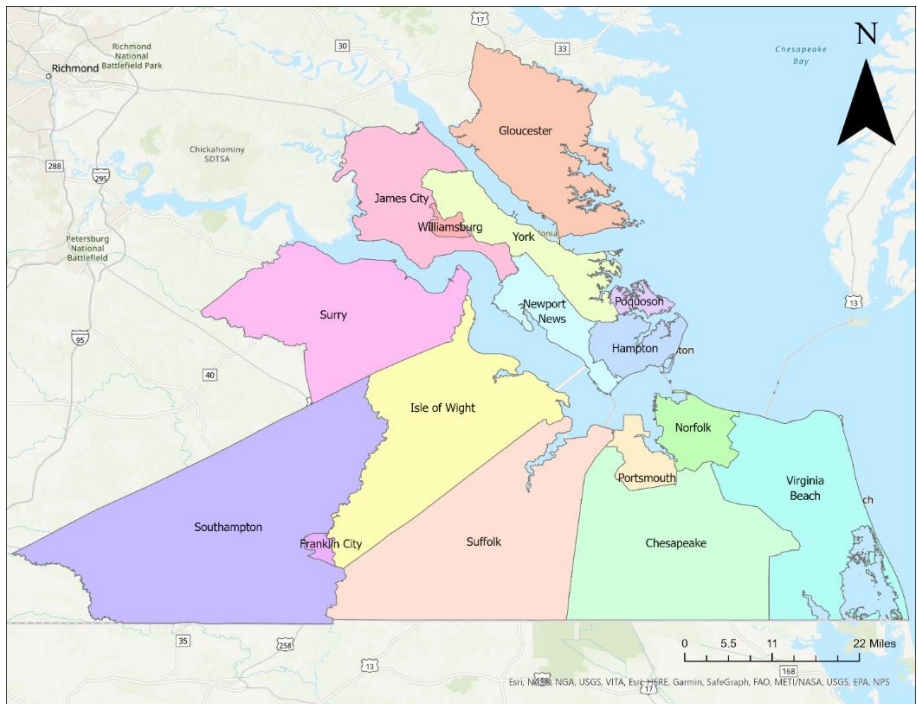

**Figure 1 Case Study Area: Hampton Roads ©Esri, NASA, NGA, USGS, VITA, Esri, HERE, Garmin, SafeGraph, FAO, METI/NASA, USGS, EPA, NPS.**

## 2. LITERATURE REVIEW

The existing literature (*13–15*, *15–25*) offers insights regarding potentially impacted populations and means for quantifying roadway flooding impacts. Many studies (*13–22*) have used simulation methods to understanding the impact of climate change on the transport system. Typically, the climate model component uses various return periods of rainfall to generate inundation scenarios to estimate the scale of flooding events, and the traffic model component quantifies how such events affect performance in terms of change in free flow speed, delay, travel time, and other mobility indicators (*26–28*). Several studies also incorporate GIS network analysis to assess the impact of flooding to the network (*13–17*), and others use numerical modeling (*18–22*) to enhance this type of analysis.

A smaller body of studies (*13–15*, *19*, *23–25*) have examined how major climate events affect accessibility, typically focusing on a single trip purpose or subgroup of trips. Pregnolato and Kasmalkar investigated work disruptions due to extreme rainfall or storm surges (*13*, *23*). Yin et al. (*14*), Shi et al. (*15*), and Coles (*19*) explored the impact of cascading effects of sea level rise and coastal flooding on emergency response. Dong et al. analyzed the effect of disaster flooding on access to health infrastructure (*24*). Stefanska et al. investigated the impact of atypical floods on grocery shopping trips (*25*). Such studies focus on large-scale climate events because disasters bring long-lasting consequences and are widely noticed.

Fewer studies have investigated the effects of recurrent flooding events. Koetse (*29*) showed that recurrent flooding has substantial impacts on travel safety where precipitation increases the frequency of crashes, but due to slower driving speeds in wet weather, injury severity decreased. Praharaj et al. (*9*) found that recurrent flooding reduces network-wide vehicle-hours of travel compared to days without such events. Using simulated higher water levels in Honolulu, Shen (*7*) found that nuisance flooding modestly reduced accessibility (on the order of 4% to 15%) for most TAZs.  However, some TAZs showed a complete or almost-complete elimination of accessibility, indicating that impacts can vary substantially by location. Except for Shen (*7*), most studies have focused on mobility rather than accessibility effects.

The socioeconomic characteristics of the population affected by climate events are important indicators for policy makers because such characteristics can help agencies direct resources where they are most needed.  The Centers for Disease Control and Prevention Social Vulnerability Index (CDC SVI) is a methodology for identifying and mapping the communities that will most likely need support before, during, and after a hazardous event (*30*). Although a few papers have investigated the relationship between social vulnerability and transportation accessibility, none specifically address the context of climate events. For example, Deboosere et al. (*31*) measured the accessibility to low-income jobs for vulnerable residents by public transport. Boisjoly et al. (*32*) quantified hospital access (based on the ability of individuals to reach a hospital but public transportation within 45 minutes and the hospital's capacity) and then related these to four characteristics:  income, unemployment, immigration status, and portion of income spent on rent. These studies connect social vulnerability with accessibility, although there is a need for further understanding of how temporal variation could affect the results.  For example, Boisjoly et al. (*32*) points out that when measuring hospital access, a 10 a.m. departure time was used to represent transit availability during a non-peak time; however, the authors specifically cite this limitation in their paper, explaining that "individuals may need to visit the hospital at any time of the day." Consideration of temporal variation becomes critical for the analysis of short-term hazard events such as recurrent flooding.

These existing studies clearly show that recurrent flooding has the potential to weaken the transportation system in coastal areas. However, they also suggest that the degree of this impact is not well understood, in part due to four gaps in the literature: reliance on simulated or forecast data (rather than observed data), a greater focus on extreme events (rather than recurring events), measurement of mobility (rather than the cumulative impact on access), and a lack of consideration for who is being impacted during specific time periods (spatial and temporal heterogeneity). Resolving these four gaps is the motivation behind this study.  In this study, crowdsourced location-specific and time-stamped flood reports and traffic congestion information are used to analyze the change in work and non-work accessibility during specific times of day due to recurrent flooding. Highly affected TAZs are identified, and then the social vulnerability of these highly impacted TAZs is analyzed.

## 3. DATA SOURCES AND PRE-PROCESSING
This section introduces the sources and the pre-processing of the data that were used in the study. The accessibility analysis incorporates data describing the transportation network, the destinations (points-of-interest), roadway flooding incidents and subsequent traffic impacts, and the socioeconomic characteristics of the residents. Maps throughout this paper were created using ArcGIS® software by Esri[1].

### 3.1 Transportation Network Data

---

[1] ArcGIS® and ArcMap™ are the intellectual property of Esri and are used herein under license. Copyright © Esri. All rights reserved. For more information about Esri® software, please visit www.esri.com.

The Hampton Roads Transportation Planning Organization (HRTPO) maintains a GIS shapefile of population and employment for each of the 1,173 TAZs in the region (*33*). Using the most recent (2015) shapefiles, the TAZ is the spatial unit of analysis.

The road network shapefile was obtained from the Virginia Geographic Information Network. The Virginia Road Centerlines file contains each road's name, direction, local speed, and number of lanes (*34*), which enables the use of ArcGIS Pro software to build the network dataset and to calculate the travel impedance for accessibility measures.

### 3.2 Crowdsourced Points-of-Interest (POI) Data

Key points-of-interest (POIs) are identified through the publicly available OpenStreetMap (OSM) (*35*). OSM provides the physical infrastructure features such as roads or buildings using tags attached to one of three types of data structures: nodes, ways, and relations. Users can obtain the locations of POIs by using the amenity key. A total of 2,263 POIs were extracted from the OSM applications programming interface (API) Overpass using Python. Figure 2 shows the locations of points-of-interest within the Hampton roads region.

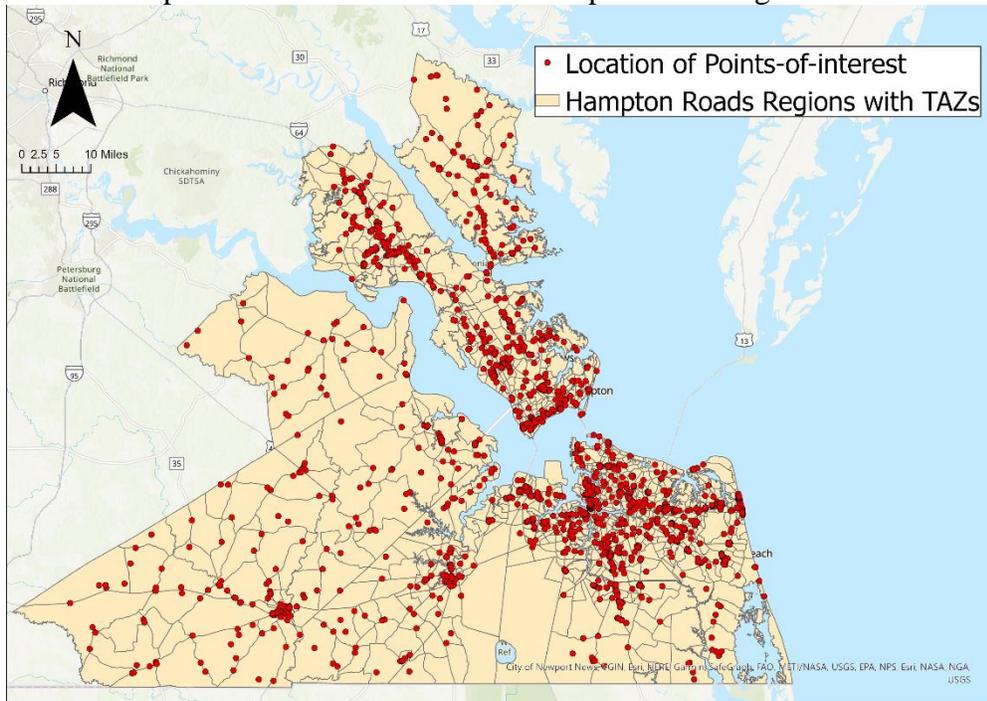

**Figure 2 Locations of Points of Interest in Hampton Roads ©Esri, VGIN, Esri, HERE, Garmin, SafeGraph, FAO, METI/NASA, USGS, EPA, NPS, Esri, NASA, NGA, USGS**

### 3.3 Crowdsourced Flood Incident and Traffic Data

WAZE, a mobile navigation application, collects flood incident reports (provided by users) and traffic condition data. The aggregated user-reported incident data are provided by WAZE as part of its Cities data-sharing program. Time-stamped and location-specific incident reports related to flooding and congestion were obtained in the Hampton Roads region for August 2018. Because crowdsourced Waze data can be subject to error or misreporting (36), traffic condition data was used for verification of flood incident reports. If a flood incident report is located within a 50-meter buffer of a "jam" location (e.g., where WAZE recorded a delay with queue), the flood report is deemed verified and the corresponding congested speed is assigned to the report location. If the flood incident report is not within 50 meters of a jam location, this unverified flood report is removed from analysis.

During August 2018, 12 days contained verified flood reports, with 8 days that had 5 or more verified reports associated with congestion data. These 8 days were chosen for analysis, as the 4 days with less than a handful of flooding reports across the entire region were believed to have negligible accessibility impacts. The flood reports within these 8 days (August 2, 10, 11, 12, 20, 21, 30, and 31) were aggregated across five time-of-day periods: period 1 (12:00 to 6:00 am), period 2 (6:00 to 9:00 am), period 3 (9:00 am to 3:00 pm), period 4 (3:00 to 6:00 pm), and period 5 (6:00 pm to 12:00 am), matching the time-of-day disaggregation in the regional travel demand model.

Table 1 contrasts the number of flood reports before and after verification for each time period across all eight days. Figure 3 shows the locations of all flood reports for all periods in red and the locations of jam data verified flood reports in green. Based on the verified flood incident reports, there are total of 20 unique date-time period combinations in the analysis, indicating that 20 unique time periods (out of 155) in August 2018 experienced significant recurring flooding. Table 1 aggregates these 20 unique date-time period combinations across different times of day, to show the temporal variation in quantities of flood incident reports.

**Table 1 Verified Flood Incident Reports per Time Period**

| Time period | # Original reports | # Reports associated with congestion data | % Reports associated with congestion data |
|---|---|---|---|
| Period 1 (12:00am to 6:00am) | 8 | 4 | 50.0% |
| Period 2 (6:00am to 9:00am) | 63 | 21 | 33.3% |
| Period 3 (9:00am to 3:00pm) | 48 | 24 | 50.0% |
| Period 4 (3:00pm to 6:00pm) | 177 | 133 | 75.1% |
| Period 5 (6:00pm to 12:00am) | 108 | 47 | 43.5% |

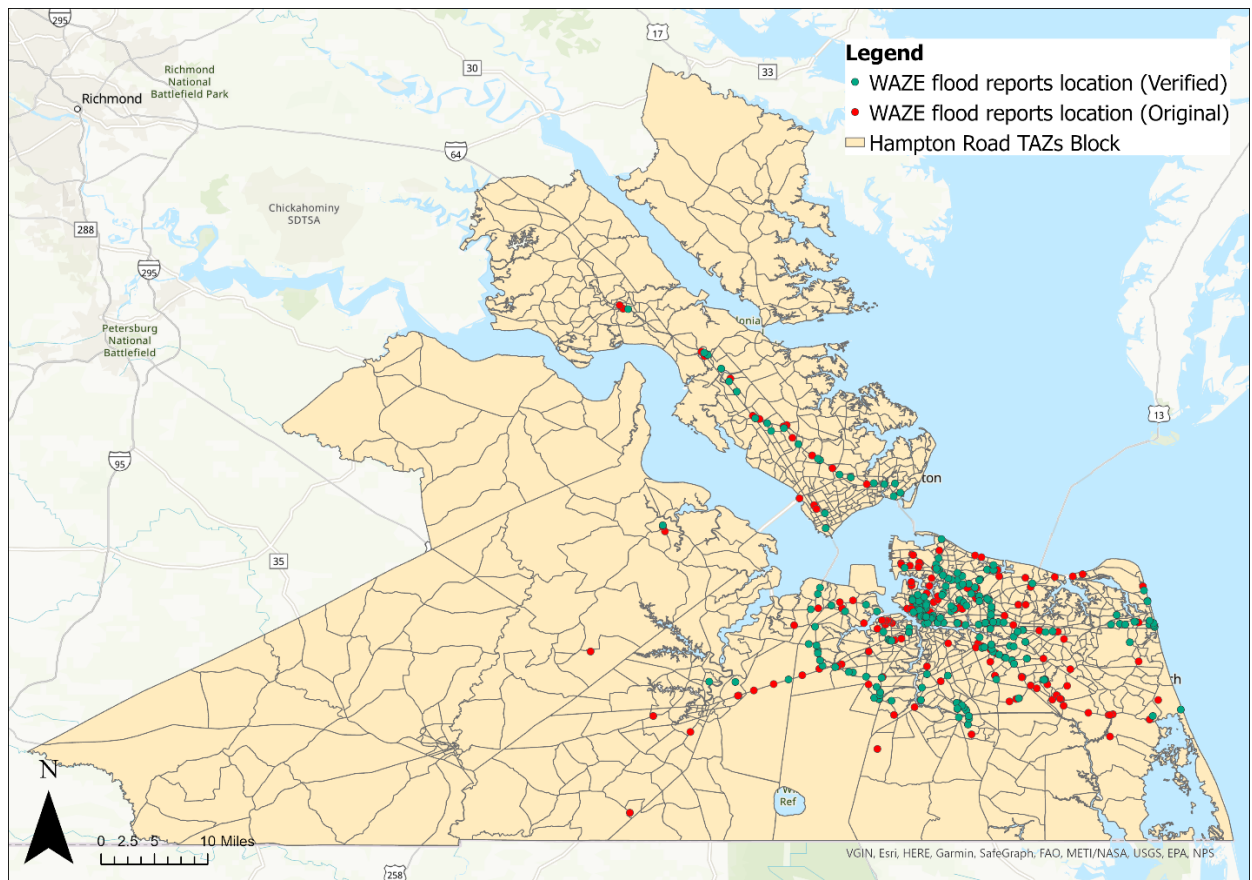

**Figure 3 Locations of WAZE flood reports. ©VGIN, Esri, HERE, Garmin, SafeGraph, FAO, METI/NASA, USGS, EPA, NPS**

### 3.4 Socioeconomic Data

Population and employment data by TAZ (*33*) were used to identify the non-zero population and non-zero employment TAZs. Census tract level socioeconomic data from the 2017 American Community Survey (*38*) were used to compute the social vulnerability indices. TAZs and census tracts boundaries do not align, thus each TAZ was associated with the census tract that contains the largest portion of the TAZ.

### 4. METHODOLOGY

Yin et al. (*3*) suggested five categories of variables for defining accessibility measures, and the framework is adapted in this study: 1) the spatial unit (TAZ); 2) the types of opportunities (employment and non-work-related POIs); 3) the mode of transportation (driving); 4) the origins and destinations (TAZs with non-zero total population and TAZs with jobs or non-work POIs, respectively); and 5) travel impedance (calculated using gravity model).

There are two types of accessibility measured in this study: work-related accessibility (based on employment) and non-work-related accessibility (based on POIs). For both measures, only TAZs with non-zero total population (1115 out of 1173 TAZs) are considered as origins. Similarly, only TAZs with non-zero employment (990 TAZs) or non-zero count of POIs (670 TAZs) are used for destinations. In order to calculate the travel time between TAZs, each of the origin TAZs was paired with every destination TAZ, resulting in an exhaustive set of origin-destination (OD) pairs in the analysis. The accessibility of each TAZ is calculated based on the zone centroid.

## 4.1 Accessibility Calculation
### 4.1.1 Model selection

It is unlikely that recurrent flooding events will completely eliminate the use of the major roadways. Rather, the impacts are felt as the sum of incremental increases in travel time. For this reason, a better measure for reflecting the cumulative impact of recurrent flooding is a comprehensive index. A Gamma-based impedance function based on a database of practice from large metropolitan planning organizations (MPOs) (*40*) was chosen to be consistent with the region's long range planning process. The long range regional model (*41*) includes a gravity-based measure to reflect travel impedance. In terms of weighing travel time, parameters consistent with Cambridge Systematics et al. (*40*) values for "MPO 1" are chosen for this study as it exhibited a medium amount of sensitivity with respect to the work trip.

### 4.1.2 Model calculation
The method of Primal Measures: Opportunity-Denominated Access (*2*) in combination with the gravity model is used to calculate accessibility in this study. This measure is known as the Hansen equation, modified to include the zone's population (Equation 1).

$$A_i = P_i \sum_j O_j \, f\big(C_{ij}\big) \tag{1}$$

Where:

$A_i$: access from TAZ $i$.

$P_i$: population of TAZ $i$.

$O_j$: number of opportunities available at destination TAZ $j$.

$C_{ij}$: cost of travel from TAZ $i$ to $j$.

$f\big(C_{ij}\big)$: impedance function, also known as friction factor.

In this study, the cost of travel from $i$ to $j$, $C_{(ij)}$, is represented by travel time, using the Gamma function (*40*) from the gravity model as shown in Equation 2.

$$F_{ij}^{p} = a * t_{ij}^{b} * exp\big(c * t_{ij}\big) \tag{2}$$

Where:

$F_{ij}^{p}$: friction factor for trip purpose $p$ from TAZ $i$ to $j$.

$t_{ij}$: the travel impedance between TAZs $i$ and $j$, travel time.

$a$: gamma function scaling factors, does not change the shape of the function.

$b$: gamma function scaling factors, always negative.

$c$: gamma function scaling factors, generally negative.

Since the parameter does not change the shape of the function and is constant, this study used $a=2$ in the calculation as suggested in the reference. The values of parameters b and c are provided by Cambridge Systematics et al. (*40*) based on the "large MPO" classification of Hampton Roads, reflecting a population of over a million people. For this study, the values of parameters for home-based work trips are b = -0.503 and c = -0.078. For home-based non-work trips, the values of parameters are b = -3.993 and c = -0.019.

### 4.1.3 Generate OD matrix and calculate change in accessibility

The travel cost for each OD pair $t_{ij}$ is generated in ArcGIS Pro 2.8 by the network analysis function, based on a local network dataset built from the roadway shapefiles obtained from HRTPO.

An OD matrix of travel times across an uncongested traffic network (operating at free flow speed) is built as base case scenario. Accessibility is calculated for the base case scenarios as the baseline for accessibility comparison with the flood disruption scenarios. The output of the OD matrix with travel time is stored in the line layer attribute table which contains the value of $t_{ij}$ from each origin TAZ to each destination TAZ. The OD matrix analysis tool in ArcGIS provides a layer named "line barriers." This layer is a feature class in network analysis that restricts access or alters costs of the underlying edges and junctions of the associated network dataset, in order to model temporary changes to the network. In this study, the locations of verified flood reports were identified as temporary changes to the network, with the associated WAZE jam record serving as input (of additional travel time cost projected onto the closest roadway network segment) to this feature class. In ArcGIS, the cost of line barriers is set to be scaled cost, which indicates how much travel cost is scaled when the barriers are traversed by vehicles. The scaling factor is ratio of local speed (from GIS layer) and jam speed (from WAZE data), then the travel time is adjusted for the flood disruption scenario. With the updated line barriers layer reflecting flood-related travel delays, the OD matrix is solved again in ArcGIS, and a new output is generated and stored in the line layer attribute table. This process is executed for both work- and non-work-related accessibility, by changing the origin and destination input to reflect the different OD pairs. R Studio is used to calculate the accessibility based on Equations (1) and (2) for each scenario.

The percentage change in accessibility from the base case to the flood disruption scenario is computed based on Equation (3) for each origin TAZ.

$$\Delta A_i = \frac{A_{b,i} - A_{f,i}}{A_{b,i}} * 100 \tag{3}$$

Where:

$\Delta A_i$: accessibility difference between base scenario and flood scenario of TAZ *i*.
$A_{b,i}$: accessibility for TAZ *i* in base scenario.
$A_{f,i}$: accessibility for TAZ *i* in flood disruption scenario.

Based on Equation (3), the greater the value of $\Delta A_i$, the greater the impact of recurrent flooding on accessibility in TAZ *i*. Note that because the population $P_i$ of a given zone does not change from the base scenario to the flood scenario, the term $P_i$ in Equation (1) does not affect the results in Equation (3).

### 4.2 Generate social vulnerability index

A social vulnerability index (SVI) was implemented for exploring the characteristics of communities whose accessibility is highly affected by recurrent flooding events based on the aforementioned CDC methodology (*30*) and following the structure provided by Masterson et al. (*43*). First order SVI variables are derived from previous studies related to climate events such as floods (*44–47*) and other natural hazards (*44*, *48*, *49*). Variables for accessibility to work and other destinations for socially vulnerable groups are selected based on the report from Gassmann et al. (*50*). Figure 4 illustrates the SVI framework used for this study.

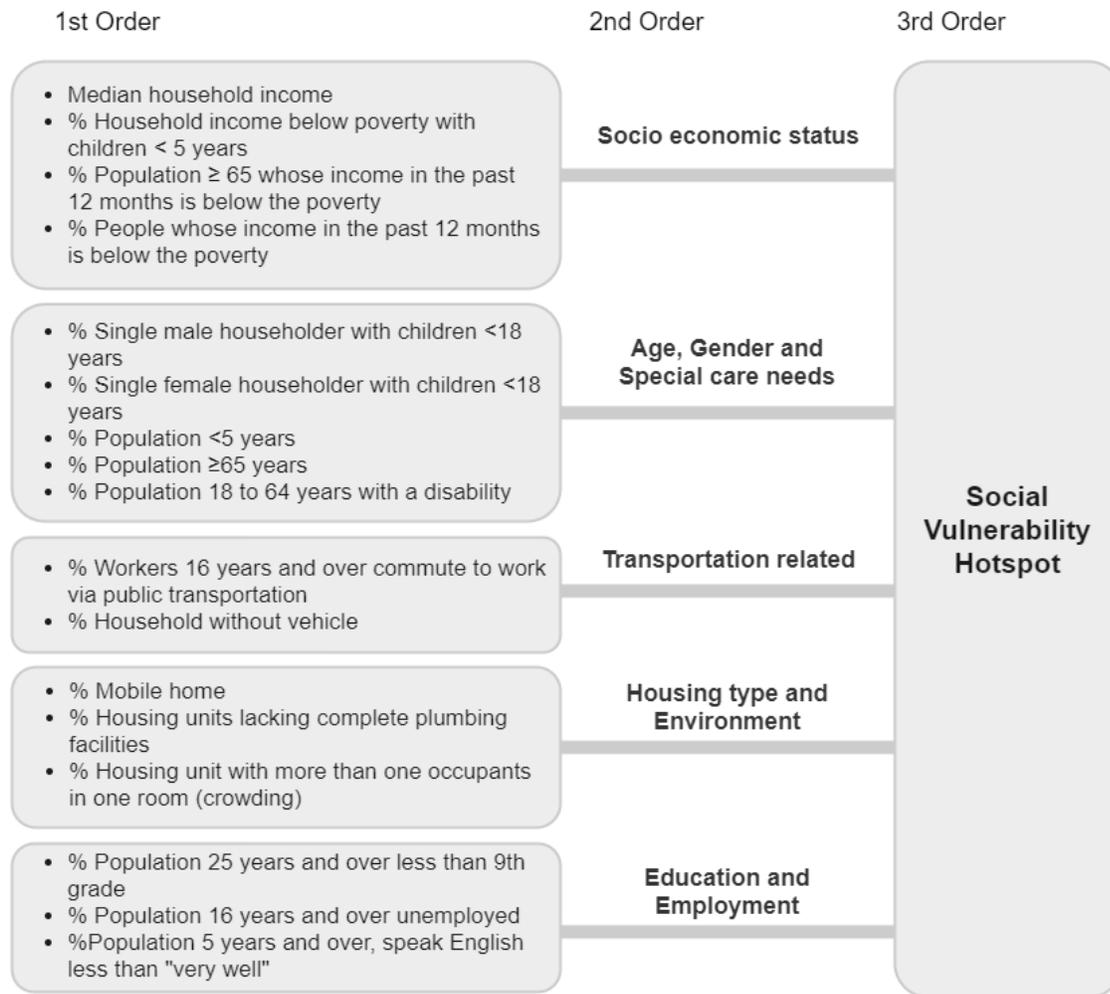

**Figure 4 Social vulnerability index (SVI) framework**

Data for the first order variables are obtained from the U.S. Census Bureau (*38*). The Hampton Roads region contains 413 census tracts, with 399 census tracts containing valid data values for computing the SVI. The remainder 14 census tracts have no sample observations or too few sample observations to compute an estimate, therefore these census tracts are excluded from the analysis. Given that there are almost three times as many TAZs as census tracts, typically each census tract contains more than one TAZ. Then, if the TAZ is completely contained in the census tract, then the socioeconomic properties of the tract can be assigned to the TAZ. 125 out of 1,173 TAZs spanned more than one census tract. In that case, the TAZ is assigned the socioeconomic properties of the census tract that contains the largest portion of the TAZ's area.

The SVI first order variables (as shown in Figure 4) of the TAZs are normalized using the min-max scaling method given in Equation 4.

$$x_{scaled} = \frac{(x - x_{min})}{(x_{max} - x_{min})} \tag{4}$$

Where:

    $x_{scaled}$ : scaled value
    $x$: original cell value
    $x_{min}$: minimum value of the first order variable
    $x_{max}$: maximum value of the first order variable

The normalized first order SVI values range from 0 to 1, with higher values indicating greater vulnerability. To compute the second order SVI, the normalized value from the first order SVI for each TAZ is summed across each of five domains shown in Figure 5: socioeconomic status; age, gender, and special care needs; transportation; housing type and environment; and education and employment. Then, the normalization procedure is once again applied to determine the second order SVI. Similarly, the third order SVI for each TAZ is calculated as the sum value of the five domains (*51*) and then normalized again to yield the final third order SVI, ranging from 0 to 1.

## 5. RESULTS AND DISCUSSION

Recurrent flooding-induced accessibility changes are assessed for 1,113 of 1,173 TAZs in the Hampton Roads region, as 58 TAZs have zero population and 2 TAZs were missing valid road network information in GIS. Figures 5(a) - 5(e) show the accessibility reduction ($\Delta A_i$) of each TAZ for work-related travel during each of the five time-of-day periods for August 2018. Similarly, Figures 6(a) - 6(e) show the $\Delta A_i$ for each TAZ for non-work-related travel. TAZs with less than 1% reduction in accessibility are considered low impact (light green). TAZs with reduction in accessibility between 1 and 5%, inclusively, are considered medium impact (medium green). Lastly, TAZs with reduction in accessibility above 5% are considered high impact (blue).

### 5.1 Accessibility Impacts for the Total Population

Geographically, the differentiation of work and non-work accessibility matters. For the morning peak (6 to 9am), work accessibility is hampered greatly in the upper peninsula region, as shown by the blue in Figure 6(b). By contrast, for non-work accessibility, the greatest impacts are experienced in parts of this peninsula region but also near the southernmost inland area as shown in Figure 7(b). These results suggest greater work accessibility impacts for James City County and greater non-work accessibility impacts for the City of Chesapeake during the morning peak travel period.

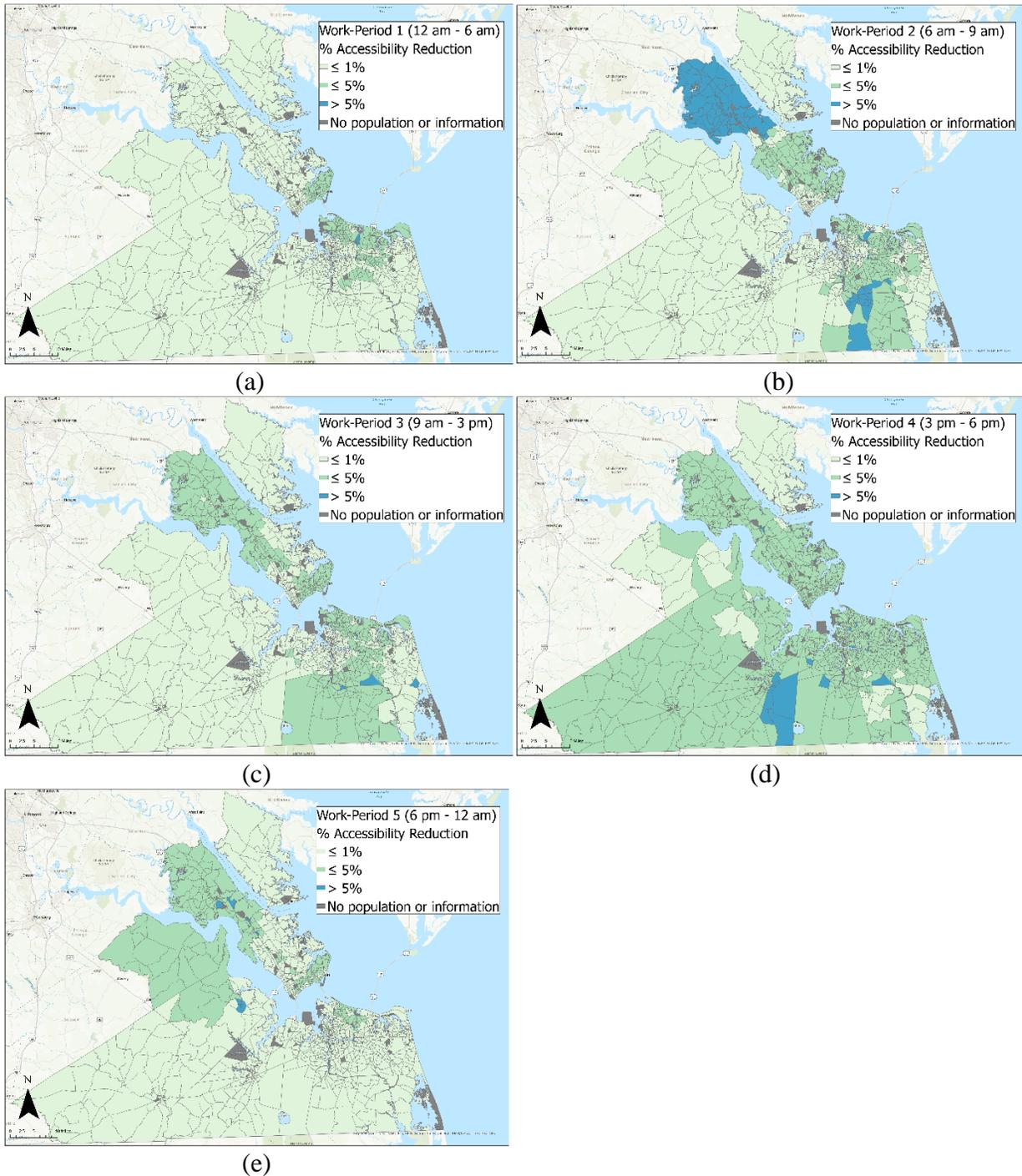

(a)

(b)

(c)

(d)

(e)

**Figure 5 Percent Change in Work Accessibility in (a) Period 1, (b) Period 2, (c) Period 3, (d) Period 4, and (e) Period 5**

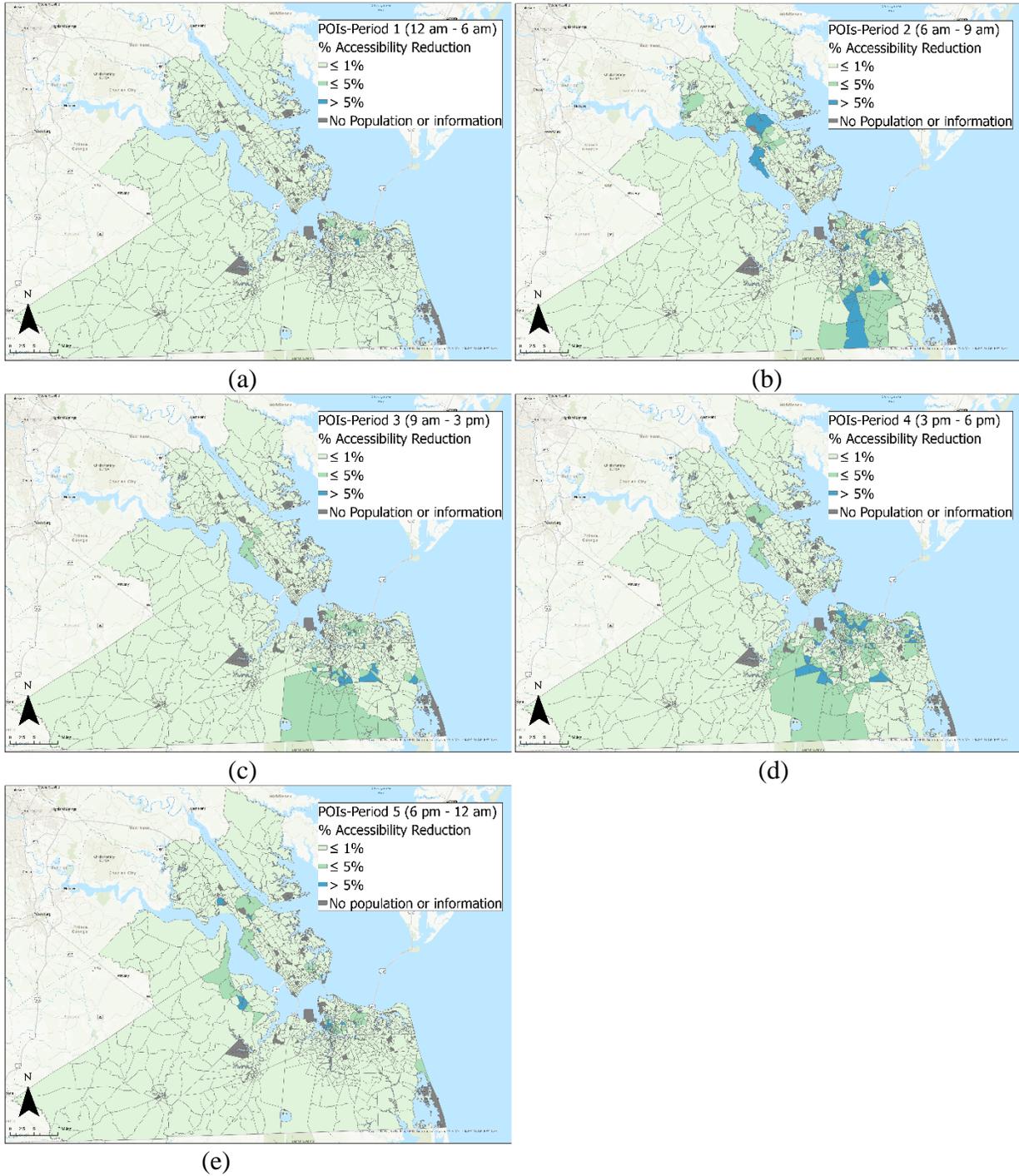

(a)                                            (b)

(c)                                            (d)

(e)

**Figure 6 Percent Change in POIs Accessibility in (a) Period 1, (b) Period 2, (c) Period 3, (d) Period 4, and (e) Period 5**

Table 2 summarizes the percentage of TAZs that fall within the three categories of accessibility impact over the five time-of-day periods for both work and non-work trips.

**Table 2 Temporal Distribution of Accessibility Impacts**

| Destination Type | Time Period [a] | % TAZs Impacted (n=1,113) | | | Maximum Accessibility Reduction in a Single TAZ ($\Delta A_i$) | Population weighted mean accessibility reduction |
|---|---|---|---|---|---|---|
| | | Low Impact (⩽1%) | Medium Impact (⩽5%) | High Impact (>5%) | | |
| **Work** | 1 (midnight to 6 am) | 81.0% | 18.5% | 0.5% | 13.8% | 0.78% |
| | 2 (6 am to 9 am) | 58.5% | 31.8% | 9.7% | 49.6% | 1.71% |
| | 3 (9 am to 3 pm) | 60.7% | 38.9% | 0.4% | 14.5% | 1.11% |
| | 4 (3 pm to 6 pm) | 6.1% | 91.9% | 2.0% | 9.7% | 2.09% |
| | 5 (6 pm to midnight) | 78.7% | 20.7% | 0.6% | 14.6% | 0.62% |
| **Non-Work** | 1 (midnight to 6 am) | 97.0% | 2.5% | 0.5% | 38.6% | 0.16% |
| | 2 (6 am to 9 am) | 88.1% | 8.4% | 3.5% | 87.9% | 0.81% |
| | 3 (9 am to 3 pm) | 91.6% | 6.8% | 1.5% | 24.9% | 0.44% |
| | 4 (3 pm to 6 pm) | 74.9% | 19.7% | 5.4% | 21.8% | 1.00% |
| | 5 (6 pm to midnight) | 91.5% | 6.1% | 2.4% | 30.0% | 0.46% |

[a] Time periods include the beginning but not ending time point: a flood event that occurs at precisely 6:00:00 a.m. is attributed to period 2 rather than period 1.

As shown in Table 2, for both work and non-work-related trips, recurrent flooding events incurred the greatest accessibility reduction in a single TAZ during Period 2, with maximum accessibility reduction of 49.6% (with population-weighted mean reduction of 1.71%) and 87.9% (with population-weighted mean reduction of 0.81%) in the most affected TAZ, respectively. Similarly, the number of TAZs with more than 5% of accessibility reduction (high impact) was the greatest for work-related trips during Period 2. However, for non-work related trips, Period 4 had the greatest number of TAZs that were highly impacted by recurrent flooding. Although the destinations and opportunities are different for work and non-work trips, the highly affected areas are relatively similar for Periods 3, 4 and 5 as shown in Figure 6 (c), (d), (e) and Figure 7 (c), (d), (e). These results suggest that the accessibility impact of recurrent flooding varies by trip purpose and time of day. Of the 1,113 TAZs, the population weighted mean accessibility reduction for work-related trips for highly impacted TAZs across all time periods was 9.72%, compared to 17.12% for non-work trips. Part of this disparity is due to work trips being less sensitive to the spatial patterns of flood incidents (and scaled travel impedance) compared to non-work trips, as the destinations for work trips are more evenly distributed across the region compared to non-work trips, whose destinations exhibit more clustering. Another contributing factor is the number of observations: because traffic volumes are greatest during the morning and evening peak periods, there are more travelers who can report flooding incidents for periods 2 and 4. In that sense, traffic volume is an indirect parameter in this study's methodology.

For work trips in August 2018, 415 unique TAZs experienced significant flood-related accessibility reduction for more than one time period, whereas for non-work tips, 128 unique TAZs experienced more than 5% accessibility reduction for more than one period. The 4 TAZs that experience the most frequent accessibility impacts (more than 7 time periods across 8 days) are all located in close proximity to bodies of water within the City of Norfolk, exacerbating roadway flooding with rainfall and high tide. Figure 7 shows the location and frequency of all TAZs that experience greater than 5% accessibility reduction (high impact) for at least one time period for both work and non-work trips. Once again, accessibility impacts for work trips are more wide-spread than for non-work trips, which are experienced in clusters of zones.

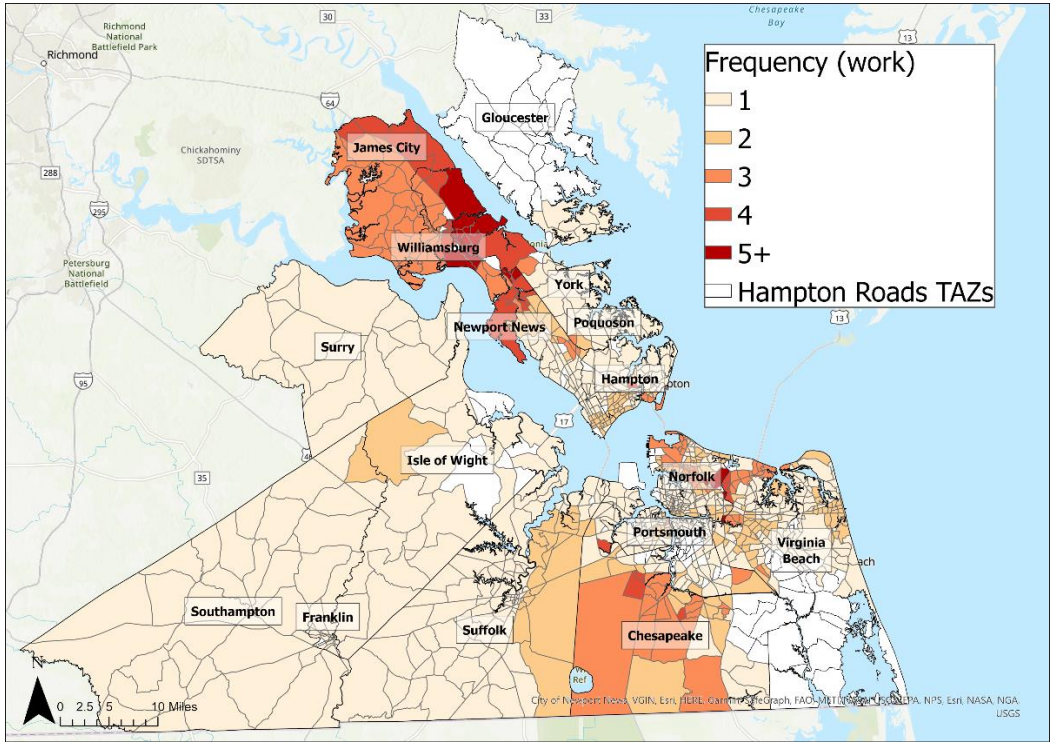

(a)

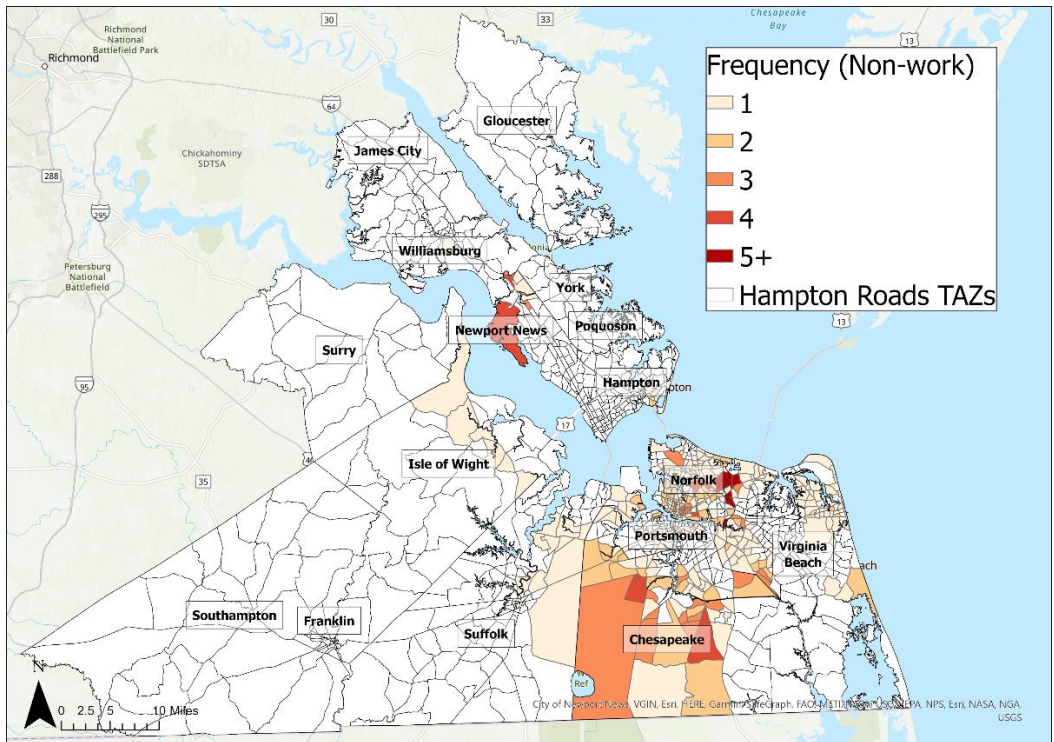

(b)

**Figure 7 Location of highly impacted TAZs with frequency for (a) work trips, (b) non-work trips. Accessibility Impacts for the Vulnerable Population**

## 5.2 Auto Accessibility Impacts for the Vulnerable Population

Principal component analysis (PCA) is applied to the second order SVI for all TAZs that are highly impacted by recurrent flooding for both trip purposes. As these variables are highly correlated with each other, compressing them into meaningful components might show more clear relationships. The relationship between second order SVI variables (socioeconomic status; age, gender, and special care needs; transportation; housing type and environment; and education and employment) and accessibility reduction for each TAZ are examined. PCA results show that for work trips, only the first two Principal components (PCs) have eigenvalue greater than 1 across 6 PCs, and accumulated amount of explained variance is 64.1%. Similarly, for non-work trips, the first two PCs have eigenvalue greater than 1 with 69.3% accumulated explained variance. Therefore, for both types of trips, only two PCs are retained and the bi-plots are shown in Figure 8.

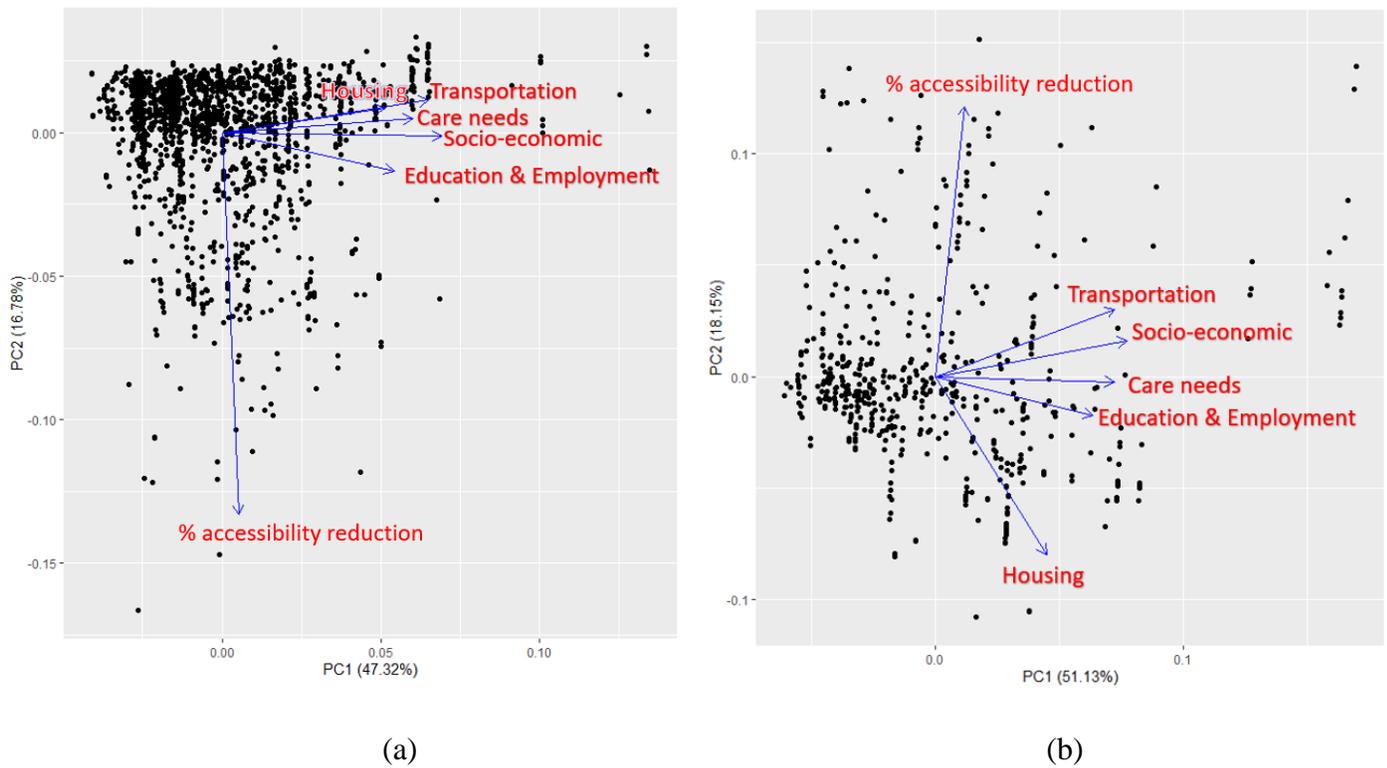

(a)                                                                                  (b)

**Figure 8 PCA biplot for (a) work trips, (b) non-work trips**

For work tips, results suggest that flood-induced accessibility reduction is positively correlated with lower socioeconomic status and vulnerability in education and employment. In practical terms, Hampton Roads residents with lower education levels, who are unemployed, and have low English language proficiency may be more prone to accessibility reduction under recurrent flooding, further exacerbating their access to employment. This result is in alignment with Hossain et al. (*52*), which suggested that people with less skilled English can experience substantially increased vulnerability to disaster flooding and after disaster recovery impacts. As this group already experiences barriers in accessing employment, the additional reduction in accessibility to work under recurrent flooding (compared to persons not in this group) is concerning. Figure 8(b) suggests that population with higher vulnerability in transportation and socio-economic status are more likely to experience accessibility reduction for non-work tips under recurrent flooding. The relationship of third order SVI with accessibility reduction was also examined with scatter plots, however the correlation was relatively weak, thus no conclusions can be drawn at this point.

## 6. CONCLUSIONS AND LIMITATIONS

Compared to extreme disaster events, the impacts of recurrent flooding have not been widely examined. In response, this study (Part 1) demonstrates a method to investigate spatial and temporal variation of accessibility impacts due to recurrent flooding using crowdsourced flood incident and congestion data. The analysis also uses a modified CDC SVI framework to analyze which vulnerable communities may be more at risk for recurrent flooding-induced transportation accessibility impacts. The analysis related to transit accessibility impacts under recurrent flooding is investigated in a second paper (Part 2).

The case study on auto accessibility impacts of recurring flooding offers two conclusions specific to the Hampton Roads region. First, the greatest accessibility reduction occurs during the morning peak period for work and non-work travel. Additionally, non-work trips experience greater accessibility reduction than work trips, due to the spatially clustered nature of non-work destinations in the region. Second, when examining the social vulnerability of highly impacted TAZs throughout the region, PCA biplots suggest that residents living in zones with higher vulnerability in education and employment experience higher accessibility reduction for work trips, and people with higher vulnerability in socio-economic status are more prone to accessibility reduction for both work and non-work trips.

This study also offers two methodological lessons that may be of interest to researchers seeking to evaluate responses to recurrent flooding in other locations.

- The study demonstrates a method to use crowdsourced empirical data to perform an accessibility analysis. While crowdsourced data are imperfect, they improve on previous work by using observed data rather than simulated data. These data facilitate a focus on recurrent flooding's temporally and spatially heterogeneous effects on the transportation system, which is distinct from static identification of zones with high recurrent flooding occurrence. In other words, a flooded location with low traffic demand degrades performance less than a flooded location with high traffic demand. This distinction allows one to prioritize scarce public resources on the basis of accessibility—that is, from a user's perspective, on the basis of the greatest improvement to transportation system performance.

- The study demonstrates how to consider vulnerable users. For work-based accessibility, the analysis showed that locations meeting certain social vulnerability characteristics can be more affected by recurrent flooding than the region at large. Such analyses can help policy makers understand transportation investments that can minimize inequitable impacts borne by these vulnerable communities, which is especially meaningful for hazards that are expected to exponentially worsen in the future.

This case study has several limitations. First, the analysis period is only one month (August was the month with the most number of flood reports for the Hampton Roads region in 2018). To protect user privacy, WAZE jam data do not match exactly to the roadway network shapefiles, requiring manual translation of the jam data line barriers in GIS. An automated method for processing the WAZE data to match the roadway network data would enable future analysis to include a longer analysis period, which would enable analysis of whether recurrent flooding impacts and social vulnerability to flooding impacts change over time. Second, the analysis is focused on auto accessibility. Socially vulnerable groups are less likely to have access to private vehicles compared to non-vulnerable groups. An improvement would be to consider the accessibility reduction due to recurrent flooding for all modes, especially public transit. Third, this study did not consider different weights for non-work POIs. For instance, a large hospital should attract more trips than a small clinic. Thus Equation 1 could be modified such that the opportunities term (e.g., $O_i$) reflects the size (or importance) of the POI. Lastly, smartphone applications (e.g. WAZE) are less likely to be adopted by marginalized and vulnerable groups (*53,54*). While a single incident report among all drivers traveling on a specific road segment or intersection is all that is needed to generate a crowdsourced report, it is possible that WAZE flood incident reports do not show all the flood-induced accessibility impacts experienced by groups less likely to use smartphone apps.

**ACKNOWLEDGMENTS**
The authors thank the Hampton Roads Transportation Planning Organization and WAZE for facilitation of data acquisition. The authors would also like to thank Shraddha Praharaj and Yige Tang for giving valuable input in reviewing the manuscript. This work is supported by the National Science Foundation's Critical Resilient Interdependent Infrastructure Systems and Processes program (Award 1735587).

**AUTHOR CONTRIBUTIONS**

The authors confirm contribution to the paper as flows: study conception and design: L. Zeng, T.D. Chen, J. Miller, F.T. Zahura, J.L. Goodall; data collection and processing: L. Zeng, F.T. Zahura; analysis and interpretation of result: L. Zeng, T.D. Chen, J. Miller; draft manuscript preparation: L. Zeng, T.D. Chen, J. Miller. All authors reviewed and approved the final version of the manuscript.